\documentclass[trackchanges]{aastex7}

\usepackage{amsmath}  
\usepackage{diagbox}  
\usepackage{threeparttable, tablefootnote}  
\graphicspath{{images/}}  

\begin{document}

\title{On the Origin of Neutron-capture Elements in r-I and r-II Stars: A Differential-abundance 
Analysis}
\shorttitle{On the Origin of Neutron-capture Elements in r-I and r-II Stars}

\author[0000-0002-0786-7307]{Pallavi Saraf}
\affiliation{Indian Institute of Astrophysics, Bangalore, India, 560034}
\altaffiliation{This work is done as part of PhD thesis of Pallavi Saraf at Indian Institute of Astrophysics.}
\affiliation{Physical Research Laboratory, Ahmedabad, India, 380059}
\email[show]{pallavi@prl.res.in}

\author[0000-0003-0891-8994]{Thirupathi Sivarani}
\affiliation{Indian Institute of Astrophysics, Bangalore, India, 560034}
\email{}

\author{Timothy C. Beers}
\affiliation{Joint Institute for Nuclear Astrophysics-Center for the Evolution of the Elements (JINA-CEE), United States of America}
\affiliation{Department of Physics and Astronomy, University of Notre Dame, Notre Dame, IN 46556, United States of America}
\email{}

\author{Yutaka Hirai}
\affiliation{Department of Community Service and Science, Tohoku University of Community Service and Science, 3-5-1 Iimoriyama, Sakata, Yamagata 998-8580, Japan}
\email{}

\author{Masaomi Tanaka}
\affiliation{Astronomical Institute, Tohoku University, Sendai, Miyagi 980-8578, Japan}
\affiliation{Division for the Establishment of Frontier Sciences, Organization for Advanced Studies, Tohoku University, Sendai 980-8577, Japan}
\email{}

\author{Carlos Allende Prieto}
\affiliation{Instituto de Astrof\'isica de Canarias, E-38200 La Laguna, Tenerife, Spain}
\affiliation{Departamento de Astrof\'isica, Universidad de La Laguna, E-38205 La Laguna, Tenerife, Spain}
\email{}


\author{Drisya Karinkuzhi}
\affiliation{Department of Physics, University of Calicut, Malappuram 673635, India}
\email{}

\begin{abstract}
We present a strictly line-by-line differential analysis of a moderately $r$-process-enhanced star ($r$-I: HD~107752) with respect to a strongly $r$-process-enhanced star ($r$-II: CS~31082-0001) to investigate the possible common origin of their heavy-element nucleosynthesis with high-precision abundances. This study employs ESO data archive high-resolution and high signal-to-noise spectra taken with the UVES (VLT) spectrograph. Considering only the lines in common in both spectra, we estimate differential abundances of 16 light/Fe-peak elements and 15 neutron-capture elements. Abundances of O, Al, Pr, Gd, Dy, Ho, Er, and detection of Tm in HD~107752 are presented for the first time. We found three distinct features in the differential-abundance pattern. Nearly equal abundances of light elements up to Zn are present for both the stars, indicating a common origin for these elements; in addition to no noticable odd-even differential pattern. In the case of neutron-capture elements, the $r$-I star exhibits mildly depleted light $r$-process elements and more depleted heavier $r$-process elements relative to $r$-II star. We also show that among $r$-I and $r$-II stars, the ratio of lighter-to-heavier $r$-process elements (e.g. [(Sr,Y,Zr)/Eu]) exhibits a decreasing trend with respect to the overall $r$-process enhancement, forming a continuous sequence from $r$-I and $r$-II stars. Finally, we discuss the necessity of multiple sites for the formation of $r$-I stars.
\end{abstract}



\section{Introduction}
\label{sec:intro}
One of the key questions in stellar archaeology is the nature of the astrophysical sites of $r$-process element production in the early Galaxy that enrich ancient low-mass metal-poor stars.
Several high-resolution spectroscopic studies of metal-poor stars from the Milky Way and its satellites have already shed some light on this topic, but the full picture remains elusive.
Understanding the production sites of the $r$-process is crucial for gaining insight into several key astronomical phenomena, including the formation of the first stars, early nucleosynthesis events, the chemical evolution of galaxies, and the processes underlying star formation.

$R$-process-rich metal-poor stars exhibit a diverse range of $r$-process enhancement, and can be divided into sub-classes:  $r$-I, $r$-II, $r$-III, and limited-$r$ \citep{Beers.Christlieb.2005, Cain.etal.2020, Holmbeck.etal.2020}. These classifications are based on the abundance of europium (Eu) in the photospheres of the stars, because Eu is mainly produced during $r$-process nucleosynthesis and is easily detectable in optical spectra. Definitions of these sub-classes are as follows \citep{Frebel.etal.2018}:\footnote{Note: The absolute abundance of element A is defined as $\log \epsilon (\rm A) = \log (\rm N_{A}/N_{H}) + 12$ and relative abundance of element A with respect to element B is defined as $[\rm A/B] = \log (\rm N_{A}/N_{B}) - \log (\rm N_{A}/N_{B})_{\odot}$, where $\rm N_{A}$ and $\rm N_{B}$ are number densities of A and B elements, respectively. The symbol ${\odot}$ represents the Solar value.}
\begin{table}[ht]
    \centering
    \begin{tabular}{ll}
        Class & Definition\\
        $r$-I &  +0.3 $<$ [Eu/Fe] $\leq$ +0.7, [Ba/Eu] $<$ 0.0\\
        $r$-II & +0.7 $<$ [Eu/Fe] $\leq$ +2.0, [Ba/Eu] $<$ 0.0\\
        $r$-III & [Eu/Fe] $>$ +2.0, [Ba/Eu] $<$ 0.0\\
        limited-$r$ & [Eu/Fe] $\leq$ +0.3, [Sr/Ba] $>$ +0.5, [Sr/Eu] $>$ 0.0\\
    \end{tabular}
\end{table}

More than one site has been proposed for the production of $r$-process elements. For example, core-collapse supernovae (CCSNe) are thought to be responsible for the production of light $r$-process elements \citep{Wanajo.etal.2011, Pinedo.etal.2012, Roberts.etal.2012, Ebinger.etal.2019}, whereas neutron star mergers (NSMs) are believed to be the primary production sites of heavy $r$-process elements \citep{Abbott.etal.2017, Abbott.etal.2019, Metzger.2017, Tanaka.etal.2017, Villar.etal.2017, Abbott.etal.2020}. The site that produces only light $r$-process elements is often referred to as the weak $r$-process site; the site that synthesizes heavy $r$-process elements is often referred to as the main $r$-process site \citep{sneden.etal.2008}. 

Several other proposed sites can potentially produce $r$-process elements \citep{cowan.etal.2021}. These include neutrino winds from CCSNe \citep{Woosley.Hoffman.1992, Takahashi.etal.1994, Arcones.etal.2007, Wanajo.etal.2018}, magneto-rotational supernovae with jets \citep{Symbalisty.etal.1985, Nishimura.etal.2006, Marek.etal.2006, Winteler.etal.2012}, collapsars \citep{Woosley.1993, Nagataki.etal.2007, Fujimoto.etal.2008, Siegel.Metzger.2018, Siegel.etal.2019}, hypernovae \citep{Nomoto.etal.2006}, and accretion disk outflows \citep{Siegel.etal.2017}. However, each of these sites has its merits and drawbacks.

Proper understanding of $r$-process nucleosynthesis depends on precise and accurate determination of elemental-abundance patterns in metal-poor stars, which in turn crucially relies on the accuracy of stellar parameters, stellar models, and atomic and molecular data. Even high-resolution and high signal-to-noise (SNR) spectra suffer from the systemetic errors mentioned above \citep[e.g., see elemental errors in][]{Cayrel.etal.2004, Arnone.etal.2005}. However, the differential-abundance analysis technique improves the precision limits in spectroscopic-abundance estimates, by (primarily) removing the errors in elemental abundances due to uncertainties in stellar parameters and transition probabilities ($\log gf$) \citep{Bedell.etal.2014, Nissen.etal.2018}.

Differential-abundance analysis is not a new technique. \cite{Danziger.1965} performed differential-abundance analyses of HD~116713 and HD~83548 relative to $\alpha$ Boo for the very first time. Later, this technique has been used by several other authors to obtain high-precision abundances and minute differences in abundance patterns \citep{Sadakane.etal.2003, Nissen.etal.2018, Ramirez.etal.2019}. Using this technique, \cite{Melendez.etal.2012} concluded that HIP 56948 could be a potential candidate in the search for Earth-like systems. This approach is now frequently used to search for planetary systems, Solar-twin stars, habitable planets, and to understand planet formation \citep{Zhao.etal.2002, Paulson.etal.2003, Huang.etal.2005, Melendez.etal.2009, Liu.etal.2014, Yana_Galarza.etal.2016, Jofre.etal.2021}. In addition, \cite{Reitermann.etal.1989} performed differential analyses for globular cluster member stars in the Small Magellanic Cloud (SMC) and the Large Magellanic Clouds (LMC), finding that the globular clusters in SMC and LMC exhibit under-abundant metals by a factor of three to four relative to the Milky Way.

Some studies have also used differential abundances to investigate metal-poor stars. \cite{O'Malley.etal.2017} performed differential-abundance calculations for metal-poor main-sequence (MS) stars with respect to the Sun. \cite{Reggiani.etal.2016} were the first to use the differential-abundance analysis technique for two extremely metal-poor ([Fe/H] $\leq -3.0$ stars, G64-12 and G64-37, both of which are carbon-enhanced metal-poor (CEMP) stars \citep{Placco.etal.2016}. In the last decade, the differential analysis of metal-poor globular clusters has also gained some attention. \cite{Koch.McWilliam.2011} studied light elements up to the Fe-peak for stars in the cluster NGC 6397. \cite{Roederer.Sneden.2011} and \cite{Yong.etal.2013} investigated differential abundances of neutron-capture elements for stars in M92 and NGC 6752, respectively. 

In this study, we performed a line-by-line differential-abundance analysis of two $r$-process-enhanced (RPE) stars with very similar stellar parameters, one a $r$-I star and the other a $r$-II star, yielding accurate abundance estimates with errors less than 0.05 dex. This precise abundance calculation permits us to consider how the sites of the natal yields from which these two classes of RPE stars may have originated.

The remainder of this paper is arranged as follows. 
Data preparation and radial velocity (RV) corrections are detailed in Section~\ref{sec:target_selection}. Section~\ref{sec:stellar_param} contains the stellar-parameter estimation of the HD~107752 with respect to a CS~31082-0001. The differential-abundance calculation and the uncertainties in abundance derivation are discussed in Section~\ref{subsec:abund_analysis}. In Section~\ref{sec:results}, we present the results derived from this study. In Section~\ref{sec:discussion}, we present the discussion and  their implications. Section~\ref{sec:conclusion} summarizes our conclusions. 
\vskip 2cm

\section{Target Selection and Data Acquisition}
\label{sec:target_selection}
We initially compiled a list of metal-poor ([Fe/H] $< -1.0$) RPE stars, along with their respective atmospheric parameters, from \citet{Gudin.etal.2021}. This sample comprises a total of 519 metal-poor RPE collected from the first four $R$-process Alliance (RPA) data releases \citep{Hansen.etal.2018, Sakari.etal.2018, Ezzeddine.etal.2020, Holmbeck.etal.2020}, JINAbase \citep{JINAbase2018}, and other recent literature. 

\begin{figure}
    \centering
    \includegraphics[width=0.8\columnwidth]{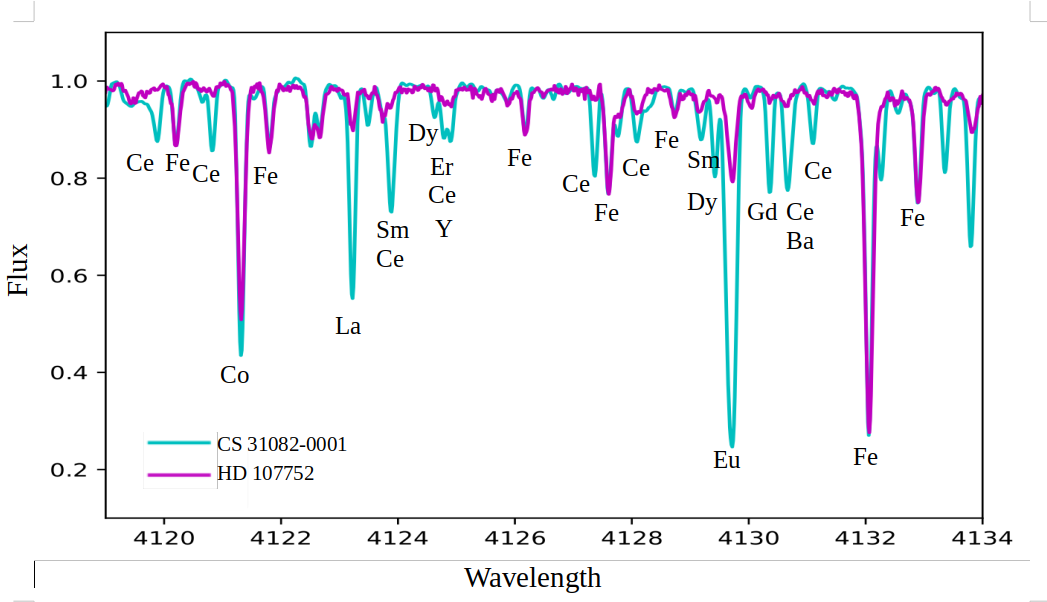}
    \caption{A comparison of stellar spectra of the $r$-I star HD~107752 (magenta spectrum) and $r$-II star CS~31082-0001 (cyan spectrum) in the Eu region. We have also marked spectral features of various elements. The enhancement of neutron-capture elements in CS~31082-0001 relative to HD~107752 is clearly evident.}
    \label{fig:comparison_Eu}
\end{figure}

Since a precise abundance comparison of $r$-I and and $r$-II stars using the differential-abundance analysis method requires high-SNR spectra of these stars having nearly identical atmospheric parameters, we searched for well-studied RPE stars in our compiled sample. We chose BPS CS~31082-0001 (hereafter CS~31082-0001), one of the most well-studied $r$-II stars. We then searched for a $r$-I star having similar atmospheric parameters; we choose HD~107752. Figure \ref{fig:comparison_Eu} shows a comparison of the spectra of these stars in the Eu region.

To reduce the errors in the analysis, it is preferable to obtain the high-resolution and high-SNR spectra of both stars from the same instrument. Thus,  we obtained the optical spectra for both stars from the European Southern Observatory (ESO) science portal \footnote{ESO Science Portal: https://archive.eso.org/scienceportal/home}. Both stars were observed with the Ultraviolet and Visual Echelle Spectrograph (UVES) \citep{UVES_Dekker.etal.2000} installed at the Very Large Telescope (VLT) facility. CS~31082-0001 was observed by R. Cayrel (Program Id: 072.D-0780 [Blue Arm], 165.N-0276 [Red Arm]); HD~107752 was observed in both the blue and red arms (at lower resolution) by C.J. Hansen (Program Id: 0104.D-0059). The basic parameters (e.g., RA, DEC, $V$ magnitude, observation date, observation ID, spectral range, spectral resolution, and heliocentric radial velocity) for the selected stars are listed in Table~\ref{tab:target_info}. Reduced spectra are available for both the stars in the ESO archive \footnote{Based on data obtained from the ESO Science Archive Facility with DOI(s): https://doi.org/10.18727/archive/50}. We have normalized the spectra using the Image Reduction and Analysis Facility (\texttt{IRAF}) package \footnote{https://iraf-community.github.io}. Next, we performed a cross-correlation of the observed spectra with synthetic spectra to calculate the geocentric radial velocities using the IDL routine \texttt{CRSCOR}. We then applied the appropriate corrections to estimate the heliocentric radial velocities. 

\begin{table*}
	\centering
	\caption{Basic Information for our program stars}
	\label{tab:target_info}
	\begin{tabular}{lccccccccc} 
		\hline
		Name & RA & DEC & $V$ mag & Obs Date & Obs ID & Range & S/N & $R$ & RV (km~sec$^{-1}$)\\
		\hline
		CS~31082-0001 & 01:29:31.13 & $-$16:00:45.49 & 11.64 & 2003-10-05 & 072.D-0780 & 3750-4978 & $>$100 & 68040 & 139\\
        CS~31082-0001 & 01:29:31.13 & $-$16:00:45.49 & 11.64 & 2001-09-05 & 165.N-0276 & 4701-6735 & $>$100 & 80930 & 139\\
		HD~107752 & 12:22:52.71 & +11:36:25.48 & 10.07 & 2020-03-03 & 0104.D-0059 & 3281-4516 & $>$100 & 40970 & 219\\
        HD~107752 & 12:22:52.71 & +11:36:25.48 & 10.07 & 2020-03-03 & 0104.D-0059 & 4617-6642 & $>$100 & 42310 & 219\\
	   \hline
	\end{tabular}
\end{table*}

We then degraded the spectral resolution of CS~31082-0001 to match that of HD~107752. For this purpose, we convolved the spectrum of CS~31082-0001 with a Gaussian whose standard deviation varies with wavelength, given by $\sigma \approx \sqrt{[(\frac{\lambda}{R_{\rm lr}})^{2} - (\frac{\lambda}{R_{\rm hr}})^{2}]}/2.355$, where $R_{\rm lr}$ and $R_{\rm hr}$ represent the spectral resolutions of the low-resolution and high-resolution spectra. CS~31082-0001 has spectral gaps of 4515-4620\,{\AA} and 5595-5765\,{\AA} and HD~107752 has spectral gaps of 4515-4620\,{\AA} and 5595-5670\,{\AA}. This results in spectral ranges 3750-4515\,{\AA}, 4620-5595\,{\AA}, and 5765-6642\,{\AA} for conducting the differential analysis.

\section{stellar Parameters} 
\label{sec:stellar_param}

To proceed further, we calculated the atmospheric parameters of the target stars. The stellar atmosphere of a star is characterized by its effective temperature ($T_{\mathrm{eff}}$), surface gravity ($\log g$), metallicity ([Fe/H]), and micro-turbulence velocity ($\xi$). Below we describe the methods used for stellar-parameter estimation.

\subsection{Line List and Equivalent Widths}
\label{subsec:linelist}
The spectroscopic method for estimating the stellar parameters requires a line list of the neutral and ionized absorption features for a species present in the spectrum. Since the optical spectrum exhibits numerous iron (Fe) lines, we made a list of Fe I and Fe II lines for both of our target stars. We consider only those lines which are not blended with other lines and have a clear continuum level. The oscillator strength ($\log gf$) and lower excitation potential (LEP) values for each line were obtained from recent literature, \citet{Heiter.etal.2015, Heiter.2020}, as used in the Gaia/ESO survey \citep{Gilmore.etal.2012}. For the detailed line list and the equivalent widths of the light elements, see Appendix~\ref{app:linelist}.

We estimated the equivalent widths (EWs) of each line using an automated code (Automatic Routine for line Equivalent widths in stellar Spectra) (\texttt{ARES}) . A detailed description of the \texttt{ARES} code is presented in \cite{ARES.2007} and \cite{ARES.2015}. \texttt{ARES} performs Gaussian fitting of the absorption lines and calculates the EWs. We also calculated the EWs manually using the \texttt{SPLOT} task in the \texttt{IRAF} package. Both methods produced similar results. For this analysis, we have accepted the lines that have EWs between 5 and 80 {m\AA}.

\begin{table*}
\centering
\caption{Atomic Data and Elemental Abundances}
\begin{threeparttable}
\begin{tabular}{ccccccc}
\hline
Element & Wavelength ({\AA}) & EP (eV) & $\log gf$ & $\log gf$ & $\log \epsilon (X)$ & $\log \epsilon (X)$\\
 & & & & Reference & CS~31082-0001 & HD~107752\\
\hline
Na I & 5889.951 & 0.000 & 0.108 & \cite{Heiter.etal.2015} & 3.719 & 3.752\\
Na I & 5895.924 & 0.000 & $-$0.144 & \cite{Heiter.etal.2015} & 3.609 & 3.724\\
Mg I & 4167.271 & 4.346 & $-$1.004 & \cite{Heiter.etal.2015} & 5.400 & 5.600\\
Mg I & 4702.991 & 4.346 & $-$0.440 & \cite{Heiter.etal.2015} & 5.250 & 5.500\\
Mg I & 5172.684 & 2.712 & $-$0.450 & \cite{Heiter.etal.2015} & 5.125 & 5.234\\
Mg I & 5183.604 & 2.712 & $-$0.239 & \cite{Heiter.etal.2015} & 5.180 & 5.213\\
Mg I & 5528.405 & 4.346 & $-$0.620 & \cite{Heiter.etal.2015} & 5.450 & 5.540\\
\hline
\end{tabular}
\begin{tablenotes}
\item [] Note: The complete table is available in machine-readable form in the online edition of the journal. A portion is shown here to illustrate its form and content.
\end{tablenotes}
\end{threeparttable}
\label{tab:line_wise_abundance}
\end{table*}

\subsection{Stellar-parameter Estimation for CS~31082-0001}
We did not directly adopt the stellar parameters for this star from the literature because we wanted to perform differential-abundance analysis; rather we performed our own analysis for estimation of its stellar parameters.
We employed the \texttt{ATLAS9} code \citep{ATLAS.1993, ATLAS.2003} for the stellar atmospheric models and 
\texttt{TURBOSPECTRUM} \citep{Turbospectrum.1998, Turbospectrum.2012} for abundance estimation.

The effective temperature ($T_{\mathrm{eff}}$) of this star is obtained by forcing a null slope of the Fe I abundance, as a function of the lower excitation potential (LEP). To estimate the micro-turbulence velocity ($\xi$), we forced a zero trend of the Fe I abundance, as a function of the reduced equivalent width, EW$_{r}=\log$(EW/$\lambda$). The surface gravity ($\log g$) was calculated assuming ionization equilibrium between neutral and ionized species, i.e., both Fe I and Fe II must provide the same abundance. The metallicity ([Fe/H]) of the star is obtained from the iron abundance for the model satisfying all three conditions simultaneously. Our analysis returns $T_{\mathrm{eff}} = 4826\ \mathrm{K}$, $\xi = 1.80~\mathrm{km~s^{-1}}$, $\log g=1.55$ (cgs), and [Fe/H] $=-2.81$ for CS~31082-0001.

\subsection{Line-by-line Differential Stellar Parameters for HD~107752}
\begin{figure}
    \centering
    \includegraphics[width=0.7\columnwidth]{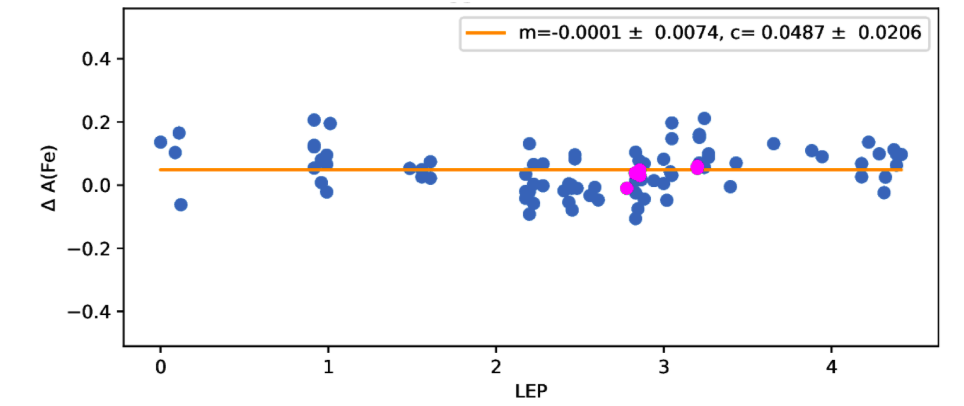}
    \includegraphics[width=0.7\columnwidth]{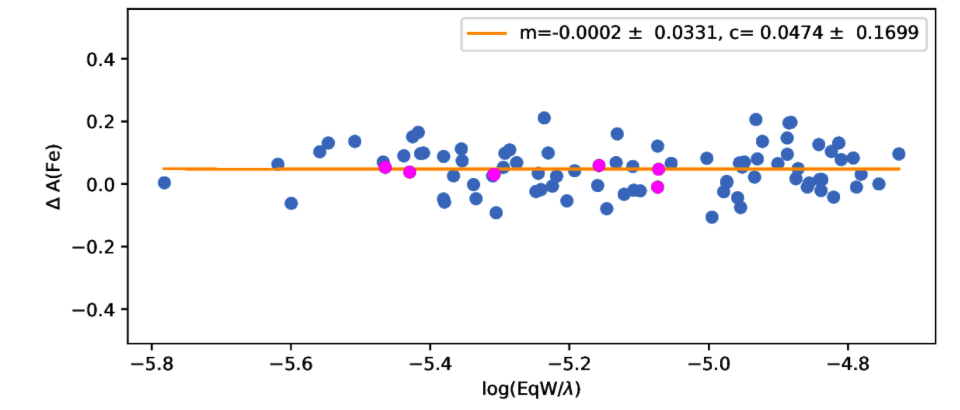}
    \caption{Stellar-parameter estimation of HD~107752 using a differential-abundance analysis with respect to CS~31082-0001. The top panel shows differential abundances of iron, as a function of lower excitation and potential (LEP), while the bottom panel shows these abundances as a function of reduced equivalent width. Blue and magenta color circles represent Fe I and Fe II features, respectively. In each panel, the gold horizontal line represents the best fit to the data. The figure legend displays the fitting parameters.}
    \label{fig:diff_stellar_param}
\end{figure}

We estimated the stellar parameters of this star using a strictly line-by-line differential approach. For this purpose, we constructed a line list of Fe I and Fe II lines that are measurable in the spectra of both this star and CS~31082-0001. 
Following the notation of \cite{Melendez.etal.2012} and \cite{Yong.etal.2013}, we define the differential abundance for $i$th line of element X as:
\begin{equation}
    \delta A_{i}^X = A_{i}^{X}(HD~107752) - A_{i}^{X}(CS~31082-0001)
    \label{eqn:diff_abund}
\end{equation}
The effective temperature ($T_{\mathrm{eff}}$
) of this star is estimated assuming the excitation equilibrium of the Fe I differential abundance:
\begin{equation}
    \frac{\partial (\delta A_{i}^{Fe\ I})}{\partial (LEP)} = 0.
    \label{eqn:teff}
\end{equation}
The micro-turbulence ($\xi$) is obtained by forcing zero trend of the Fe I differential abundance as a function of reduced equivalent width:
\begin{equation}
    \frac{\partial (\delta A_{i}^{Fe\ I})}{\partial (EW_{r})} = 0.
    \label{eqn:micro_turb}
\end{equation}
The surface gravity ($\log g$) is derived assuming ionization equilibrium between neutral and ionized species:
\begin{equation}
    \langle \delta A_{i}^{Fe\ I} \rangle - \langle \delta A_{i}^{Fe\ II} \rangle = \Delta A^{Fe\ I} - \Delta A^{Fe\ II} = 0,
    \label{eqn:logg}
\end{equation}
where $\Delta A^{X}$ represents the average abundance, $\langle \delta A_{i}^{X} \rangle$, of element X. The metallicity
([Fe/H]) estimate is taken from the metallicity of the model atmosphere. Figure~\ref{fig:diff_stellar_param} shows the stellar-parameter estimation of HD~107752 using the differential technique. Our differential analysis yields $T_{\mathrm{eff}} = 4926\ \mathrm{K}$, $\xi = 1.83~\mathrm{km~s^{-1}}$, $\log g=1.46$ (cgs), and [Fe/H] $=-2.77$ for HD~107752. Table~\ref{tab:stellar_param} lists our estimated stellar parameters for both CS~31082-0001 and HD~107752.

\begin{table}
	\centering
	\caption{Stellar Parameters for CS~31082-0001 and HD~107752}
	\label{tab:stellar_param}
	\begin{tabular}{ccccc} 
	\hline
	$T_{\rm eff}$ & $\log g$ & [Fe/H] & $\xi$ & Reference\\
	\hline
	CS~31082-0001 & & & &\\
	\hline
	4826 & 1.55 & $-$2.81 & 1.80 & This Work\\
    4825 & 1.50 & $-$2.90 & 1.80 & \cite{Hill.etal.2002}\\
    4825 & 1.80 & $-$2.91 & 1.50 & \cite{Cayrel.etal.2004}\\
     4922 & 1.90 & $-$2.78 & 1.88 & \cite{Barklem.etal.2005}\\
    4640 & 1.25 & $-$3.00 & 2.25 & \cite{Frebel.etal.2013}\\
    4846 & 1.70 & $-$2.82 & 2.25 & \cite{Frebel.etal.2013}\\
    4866 & 1.66 & $-$2.75 & 1.40 & \cite{Yong.etal.2013}\\
    4925 & 1.51 & $-$2.81 & 1.40 & \cite{Hansen.etal.2013}\\
    4650 & 1.05 & $-$3.03 & 1.55 & \cite{Roederer.etal.2014}\\
    4876 & 1.80 & $-$2.81 & 2.13 & \cite{Hansen.etal.2018}\\
    %
    %
    \hline
    HD~107752 & & & &\\
	\hline
	4926 & 1.46 & $-$2.77 & 1.83 & This Work\\
    4750 & 0.80 & $-$2.60 & 2.70 & \cite{Luck.Bond.1985}\\
    4700 & 1.70 & $-$2.69 & 1.40 & \cite{Burris.etal.2000}\\
    4370 & 0.54 & $-$3.16 & 1.58 & \cite{Zhang.etal.2009}\\
    4826 & 1.61 & $-$2.77 & 1.85 & \cite{Ishigaki.etal.2012}\\
    4826 & 1.60 & $-$2.78 & 1.90 & \cite{Ishigaki.etal.2013}\\
    4649 & 1.60 & $-$2.78 & 2.00 & \cite{Ishigaki.etal.2013}\\
    \hline
	\end{tabular}
\end{table}

\begin{figure}
    \centering
    \includegraphics[width=\columnwidth]{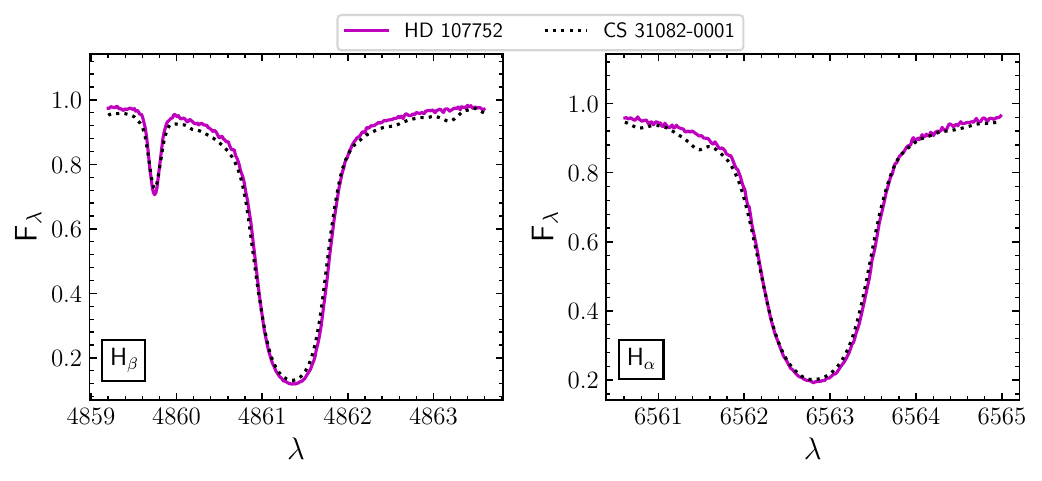}
    \caption{A comparison of stellar spectra in the H$_{\beta}$ (left panel) and H$_{\alpha}$ (right panel) regions for HD 107752 (solid magenta) and CS 31082-0001 (dotted black). The very similar profiles indicate similar temperatures for both stars.}
    \label{fig:comparison_H_profiles}
\end{figure}

We compiled the reported stellar parameters of CS 31082-0001 and HD 107752 from several past works, and list them in Table~\ref{tab:stellar_param}, along with our own measurements. For CS 31082-0001, our $T_{\mathrm{eff}}$
 estimation is close to \cite{Hill.etal.2002} and \cite{Frebel.etal.2013}; $\log g$ estimation is close to \cite{Hansen.etal.2013} and \cite{Yong.etal.2013}; [Fe/H] estimation is close to all of the studies; $\xi$ estimation is close to \cite{Hill.etal.2002} and \cite{Barklem.etal.2005}. For HD 107752, our $T_{\mathrm{eff}}$
 measurement is a bit higher than previous works; our $\log g$, [Fe/H], and $\xi$ measurements are close to \cite{Ishigaki.etal.2012} AND \cite{Ishigaki.etal.2013}. In conclusion, our stellar parameters show good consistency with previous work. Additionally, we show the comparison of H$_{\beta}$ and H$_{\alpha}$ absorption line profiles in the left and right panels of Figure~\ref{fig:comparison_H_profiles}, respectively. The wings of these profiles are sensitive to atmospheric temperature. The similar H$_{\beta}$ and H$_{\alpha}$ features observed in the two stars indicate that they have comparable effective temperatures.

\subsection{Errors in the Stellar Parameters}
To estimate the uncertainty in the calculation of stellar parameters, we have adopted the method described in \cite{Yong.etal.2013}. Uncertainty in the effective temperature ($\sigma$$T_{\mathrm{eff}}$) is derived from the 1$\sigma$ error in the slope between $\delta A_{i}^{X}$ and LEP. We varied $T_{\mathrm{eff}}$
 of the model until we obtain a new slope equal to 1$\sigma$ error value (here 0.0074). The difference between the two model temperatures provides the error in $T_{\mathrm{eff}}$
. Similarly, the uncertainty in micro-turbulence ($\sigma \xi$) is obtained from the error in the slope between $\delta A_{i}^{X}$ and EW$_{r}$. We changed $\xi$ of the model until the new slope is equal to the error in the original slope (here 0.0331). The difference between new and original micro-turbulence is the error in $\xi$. For the uncertainty in surface gravity ($\sigma \log g$), we altered the $\log g$ until we obtain $\Delta A^{Fe\ I} - \Delta A^{Fe\ II}$ equal to the total error of $\Delta A^{Fe\ I}$ and $\Delta A^{Fe\ II}$ in quadrature (i.e., $\sqrt{(\sigma \Delta A^{Fe\ I})^2 + (\sigma \Delta A^{Fe\ II})^2}$). The uncertainty in metallicity ($\sigma$ [Fe/H]) is the error in  $\Delta A^{Fe\ I}$  (i.e., the standard deviation). For HD~107752 we obtained ($\sigma$$T_{\mathrm{eff}}$
, $\sigma \xi$, $\sigma \log g$, $\sigma$ [Fe/H]) = (28 K, 0.08 km s$^{-1}$, 0.017 dex, 0.09 dex).

\section{Abundance Analysis}
\label{subsec:abund_analysis}
In the above section, we estimated the stellar parameters of CS~31082-0001 using the traditional spectroscopic method and for HD~107752 using a line-by-line differential-abundance approach.  In order to calculate the abundances of other elements by a strictly line-by-line differential-abundance estimation, we consider only lines measured in the two stars. As discussed in Section~\ref{sec:stellar_param}, we have again employed the \texttt{ATLAS9} atmospheric models for estimating the stellar parameters with the spectral synthesis code \texttt{TURBOSPECTRUM}. Both the EW analysis and spectral synthesis were performed. The EW analysis was carried out only for lines that are clear and free from blending (mainly light elements). Spectral synthesis is employed for the lines that suffer from blending (primarily heavy elements).
We also take into account hyperfine splitting as needed. In this analysis, we have utilized the Solar abundances from \cite{Asplund.etal.2009}.

First, we derived the absolute abundances of CS~31082-0001 and HD~107752 for each line identified in both stars. Using equation~(\ref{eqn:diff_abund}), we performed a strictly line-by-line differential-abundance analysis of HD~107752 with respect to CS~31082-0001. If more than one line was available for a given species, we take the average of the differential abundances as the adopted abundance for that species.

\subsection{Differential Abundances of Light and Fe-peak Elements}
We performed spectral synthesis of the CH $G$-band near 4313\,{\AA} to evaluate the differential abundance of carbon. 
Due to the high SNR of our high-resolution spectra, we are able to identify unblended lines for several light and Fe-peak elements. We first measured the EWs of these lines to estimate elemental abundances for both stars. Then, a line-by-line differential-abundance analysis is performed for each line. Among the light, odd-Z, and $\alpha$-elements, we could determine the differential abundances for Na, Mg, Al, Si, Ca, and Ti. In the iron-peak group of elements, we estimated differential abundances of Sc, V, Cr, Mn, Co, Ni, and Zn.

\subsection{Differential Abundances of Neutron-capture Elements}
For the abundance estimates of neutron-capture elements, we employed both EW analysis (for clean, blend-free lines) and spectral-synthesis methods (for blended lines). Whenever required, we have taken into account the hyperfine structure from \cite{McWilliam.1998}. We could measure the abundances of Sr, Y, and Zr among the light neutron-capture elements (first $r$-process peak elements), and Ba, Ce, Pr, Nd, Sm, Eu, Gd, Dy, Er, and Tm from the main heavy neutron-capture elements (second $r$-process peak elements). We performed a line-by-line differential abundance analysis for a total of 13 neutron-capture elements.

\subsection{Errors in the Elemental-abundance Analysis}
The total error in the abundance analysis comprises random and propagated errors. The random error comes particularly from the uncertainty in the EW measurement, oscillator strength, and the contribution of blends. The propagated error appears due to uncertainty in stellar-parameter estimation. The sum of all the errors in quadrature returns the total uncertainty in the abundance estimate:

\begin{equation}
\begin{split}
    \sigma^{2}_{\log \epsilon} &= \sigma^{2}_{\rm rand} + \sigma^{2}_{\rm prop} \\
    \sigma^{2}_{\log \epsilon} &= \sigma^{2}_{\rm rand} + \left(\frac{\partial \log \epsilon}{\partial T}\right)^{2} \sigma^{2}_{T} + \left(\frac{\partial \log \epsilon}{\partial \log g}\right)^{2} \sigma^{2}_{\log g}
     + \left(\frac{\partial \log \epsilon}{\partial \xi}\right)^{2} \sigma^{2}_{\xi} + \left(\frac{\partial \log \epsilon}{\partial [Fe/H]}\right)^{2} \sigma^{2}_{[Fe/H]}
\end{split}
\label{eqn:abund_err}
\end{equation}

where $\sigma^{2}_{\rm rand}$ and $\sigma^{2}_{\rm prop}$ are the random and propagated errors, respectively. If we use N lines to measure the abundance of species, its random error is defined as $\sigma/\sqrt{N}$. To estimate the partial derivatives of equation~(\ref{eqn:abund_err}), we changed the corresponding stellar parameter by its respective error and calculated the average of $\Delta \log \epsilon / \Delta parameter$.

\begin{table*}
	\centering
	\caption{Detailed Differential Abundances of HD~107752 with respect to CS~31082-0001}
	\label{tab:abundances_HD_107752}
	\begin{tabular}{lccccccccr} 
		\hline
Element&Species&Atomic No.&N$_{lines}$&A$^{X}$ (Solar)& mean(CS~31082-0001) &$\Delta$A$^{X}$&$\Delta$[X/Fe]&$\sigma$&\\
C&CH $G$-band&6&1&8.39&5.800&+0.090&+0.050&0.038&\\
O&O I&8&1&8.66&6.670 &$-$0.200&$-$0.240&0.022&\\
Na&Na I&11&2&6.17&3.664&+0.074&+0.034&0.069&\\
Mg&Mg I&12&5&7.53&5.281&+0.118&+0.078&0.056&\\
Al&Al I&13&2&6.37&2.940&$-$0.100&$-$0.140&0.054&\\
Si&Si I&14&1&7.51&4.680&+0.150&+0.110&0.033&\\
Ca&Ca I&20&7&6.31&3.873&+0.023&$-$0.017&0.026&\\
Sc&Sc II&21&4&3.17&+0.285&$-$0.062&$-$0.102&0.045&\\
Ti&Ti I&22&8&4.90&2.382&+0.025&$-$0.015&0.040&\\
Ti&Ti II&22&12&4.90&2.428&$-$0.065&$-$0.105&0.024&\\
V&V II&23&2&4.00&1.280&$-$0.045&$-$0.085&0.052&\\
Cr&Cr I&24&4&5.64&2.565&+0.168&+0.128&0.037&\\
Cr&Cr II&24&2&5.64&3.122&$-$0.185&$-$0.225&0.063&\\
Mn&Mn I&25&4&5.39&2.155&+0.005&$-$0.035&0.041&\\
Fe&Fe I&26&87&7.45&$-$2.810&+0.040&0.000&0.006&\\
Fe&Fe II&26&8&7.45&$-$2.810&+0.040&0.000&0.010&\\
Co&Co I&27&1&4.92&2.430&+0.050&+0.010&0.053&\\
Ni&Ni I&28&4&6.23&3.360&$-$0.111&$-$0.151&0.032&\\
Zn&Zn I&30&2&4.60&2.000&$-$0.065&$-$0.105&0.021&\\
Sr&Sr II&38&2&2.92&0.745&$-$0.405&$-$0.445&0.079&\\
Y&Y II&39&7&2.21&$-$0.161&$-$0.616&$-$0.656&0.030&\\
Zr&Zr II&40&7&2.58&0.551&$-$0.493&$-$0.533&0.042&\\
Ba&Ba II&56&4&2.17&0.325&$-$0.980&$-$1.020&0.060&\\
La&La II&57&5&1.13&$-$0.460&$-$1.120&$-$1.160&0.045&\\
Ce&Ce II&58&2&1.70&$-$0.285&$-$0.825&$-$0.865&0.034&\\
Pr&Pr II&59&2&0.58&$-$0.860&$-$0.855&$-$0.895&0.075&\\
Nd&Nd II&60&6&1.45&$-$0.087&$-$1.068&$-$1.108&0.028&\\
Sm&Sm II&62&2&1.00&$-$0.435&$-$1.090&$-$1.130&0.055&\\
Eu&Eu II&63&2&0.52&$-$0.760&$-$1.060&$-$1.100&0.023&\\
Gd&Gd II&64&3&1.11&$-$0.253&$-$0.910&$-$0.950&0.035&\\
Dy&Dy II&66&1&1.14&$-$0.060&$-$1.060&$-$1.100&0.024&\\
Ho&Ho II&67&1&0.51&$-$1.000&$-$0.930&$-$0.970&0.052&\\
Er&Er II&68&1&0.93&$-$0.320&$-$1.190&$-$1.230&0.049&\\
Tm&Tm II&69&1&0.00&$-$1.210 &$-$1.030&$-$1.070&0.047&\\
	   \hline
	\end{tabular}
\end{table*}

\section{Results}
\label{sec:results}

\subsection{Abundance Pattern of Elements up to Zn}

\begin{figure*}
    \centering
    \includegraphics[width=\textwidth]{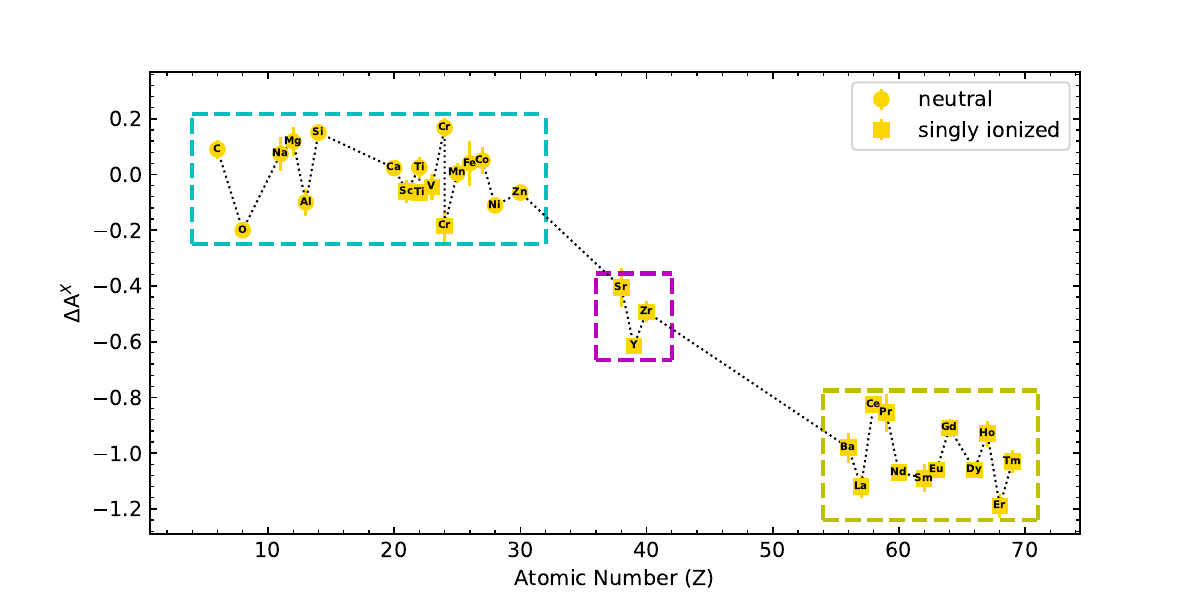}
    \caption{Differential-abundance pattern of the $r$-I star HD~107752 with respect to the $r$-II star CS~31082-0001. The circles and boxes represent neutral and ionized species, respectively. For easy identification, we have also marked each symbol with the respective atomic symbols. Elements up to Zn exhibit approximately zero differential abundance. The first $r$-process peak elements show higher differential abundances than the main $r$-process peak elements.}
    \label{fig:diff_abundance_pattern}
\end{figure*}

Figure~\ref{fig:diff_abundance_pattern} shows the differential abundances of elements as a function of their atomic numbers. The differential-abundance pattern of HD~107752 with respect to CS~31082-0001 exhibits a small excess of carbon ($\Delta A^{C} = +0.09$) in HD~107752, comparable to the iron excess ($\Delta A^{Fe} = +0.04$) within uncertainties. The carbon abundance in the lower atmospheric layers deplete during the giant phase as a result of C to N conversion during the CN-cycle \citep{Placco.etal.2014}. Thus, the carbon abundance of the natal ISM is obtained by correcting the present carbon abundance for the effects of stellar evolution. This correction mainly depends on the metallicity and initial C and N abundances. \cite{Placco.etal.2018} reported a +0.32 dex carbon correction for CS~31082-0001. Apart from the carbon abundances, both stars have very similar stellar parameters, indicating that they are at similar stages of stellar evolution. 

\cite{Yana_Galarza.etal.2021} used the differential-abundance pattern of a Solar twin to infer the mass of explosive progenitor and birth ISM metallicity. They found a clear odd-even effect in the differential pattern of the light elements, suggesting different progenitors for the Sun and Solar twin. Our differential pattern does not exhibit a strong odd-even effect, other than a very small one for Na, Mg, Al, and Si. This odd-even pattern is not present for iron-peak elements.

The $\alpha$-elements (Mg, Si, Ca, Ti) and the iron-peak (Cr, Mn, Fe, Ni, Co, Ni, Zn) elements exhibit more or less constant differential abundances (also see Figure~\ref{fig:light_lines_comparison} in Appendix~\ref{app:line_comparison} for a comparison of spectral lines). The mean differential abundances of $\alpha$-elements and iron-peak elements are 
$+0.05$ dex and $-0.01$ dex, respectively. Given the uncertainty limits, these differential abundances are equivalent to the metallicity difference of the two stars. The constant difference in the abundances suggests that these elements may have come from similar nucleosynthesis sites. The primary source of $\alpha$-elements is CCSNe \citep[see, e.g., ][]{Kobayashi.etal.2006}. However, the primary iron-peak production site is type-Ia supernovae \citep[see, e.g.,][]{Kobayashi.Nomoto.2009}. Type-Ia supernovae dominantly contribute at higher metallicity, whereas CCSNe contribute at lower metallicity. If the $\alpha$-elements and iron-peak elements in these two stars have different sources of origin, the observed constant abundance difference would be unexpected. Given that both stars are very metal-poor ([Fe/H] $\leq -2.0$), a plausible explanation for the same differential abundances of $\alpha$-elements and iron-peak elements is their similar progenitor sites, e.g., energetic CCSNe \citep{Andrews.etal.2020}. In the early Galaxy, CCSNe were likely the only source for the production of iron-peak elements. Type-Ia supernovae mainly contribute in the later stages of Galactic chemical evolution when [Fe/H] $>-1.0$ \citep{Kobayashi.etal.2011, Kobayashi.2016, Vincenzo.etal.2018}.

\subsection{Abundance Pattern of Neutron-capture Elements}
Both stars we considered are RPE stars; HD~107752 is a $r$-I star and CS~31082-0001 is a $r$-II star. Therefore, as expected, the differential abundance pattern of HD~107752 relative to CS~31082-0001 exhibits a deficiency of neutron-capture elements in HD~107752 (also see Figure~\ref{fig:heavy_lines_comparison} in Appendix~\ref{app:line_comparison} for a comparison of spectral lines). However, the differential abundances of the light neutron-capture elements (Sr, Y, and Zr) are larger than that of the heavy neutron-capture elements (from Ba to Tm). Such a large enhancement of first-peak elements relative to main-peak elements ($\approx +0.5$ dex) is generally observed in limited-$r$ stars. Moreover, $r$-I stars exhibit scatter in the first-peak element abundances, suggesting that HD~107752 represents a case of first-peak element enrichment within an $r$-I star. As we show in Section~\ref{sec:mean_r1_r1_diff_pattern}, such enrichment of the first-peak elements is typical for stars in the $r$-I sub-class.

Since both the stars are very metal-poor ([Fe/H] $\leq -2.0$) and have [Ba/Eu] $< 0.0$, we do not expect any significant contribution from $s$-process nucleosynthesis in the observed abundances of neutron-capture elements in these stars. Low- and intermediate-mass asymptotic giant branch (AGB) stars are the primary sites for $s$-process nucleosynthesis \citep{Weigert.etal.1966, Schwarzschild.Harm.1967}, which primarily contribute when [Fe/H] $> -2.0$ \citep{Truran.etal.2002}. If we assume similar astrophysical conditions are responsible for the production of $r$-process material, we anticipate the same differential abundances for light and heavy neutron-capture elements. However, we have found a significant difference between the differential abundances of the first- and second-peak elements.

\subsection{On the Uniqueness of the Differential Pattern}
\label{sec:mean_r1_r1_diff_pattern}

\begin{figure*}
    \centering
    \includegraphics[width=0.9\textwidth]{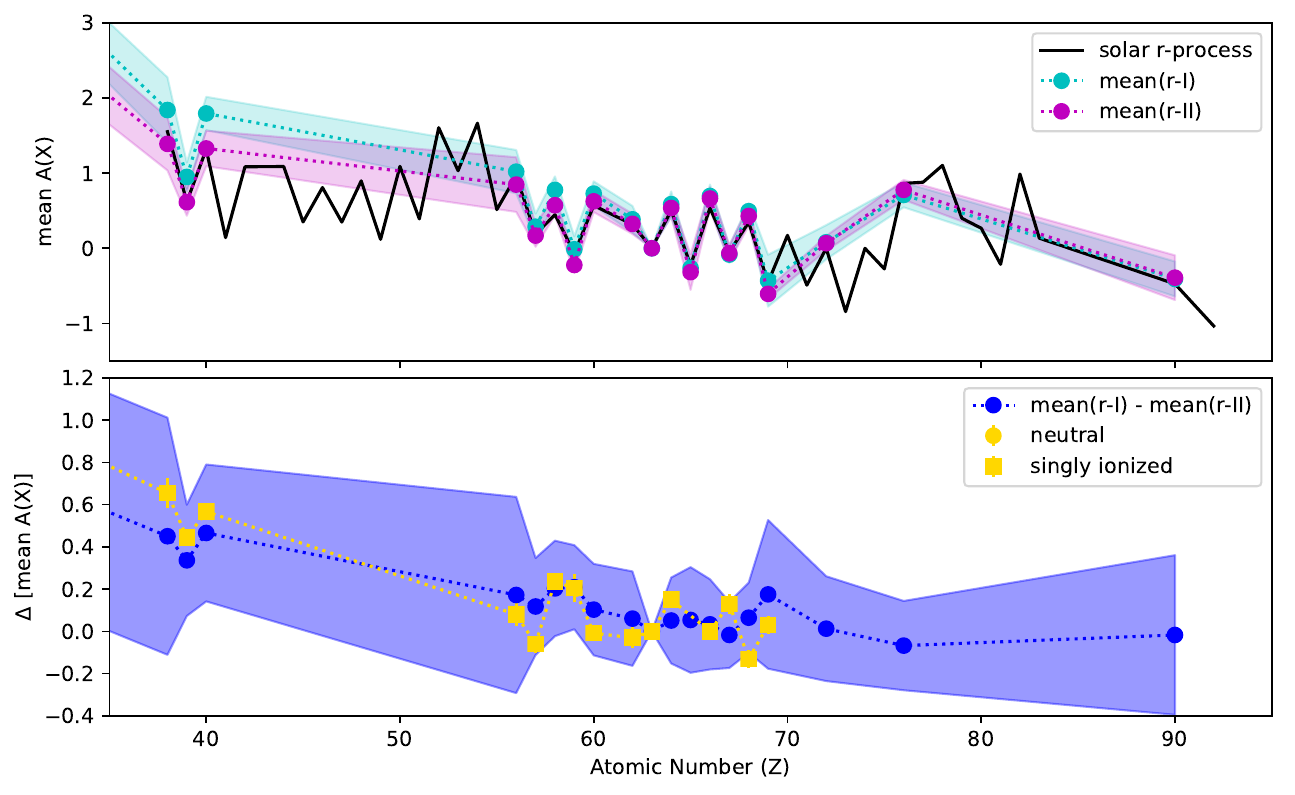}
    \caption{\textit{Top panel:} The mean abundance patterns of $r$-I (cyan) and $r$-II (magenta) stars compiled from the literature with [Fe/H] $< -2.0$. The mean of $r$-I and $r$-II are calculated after scaling the Eu of the stars to the Solar Eu. For comparison, the black curve shows the Solar $r$-process pattern. \textit{Bottom panel:} The mean differential-abundance pattern of $r$-I stars with respect to $r$-II stars in blue colored symbols; the blue shaded region shows the $1\sigma$ error range. For comparison, we have over-plotted the differential-abundance pattern of HD~107752 with respect to CS~31082-0001 using yellow colored symbols. We note a decreasing 
    differential-abundance patterm with increasing atomic number.}
    \label{fig:mean_diff_abundance_pattern}
\end{figure*}

From inspection of Figure~\ref{fig:diff_abundance_pattern}, the differential-abundance pattern of neutron-capture elements does not exhibit the same difference for first and second $r$-process peaks elements even at [Fe/H] $\approx -2.8$, when the $r$-process was the only source for the production of neutron-capture elements. Apart from the non-universal nucleosynthesis of the first and second $r$-process peak elements, we can also observe the non-constant differential abundances of elements in individual $r$-process peaks. The elements of the second peak are of particular interest, as they usually show the same pattern for  metal-poor RPE stars. To test whether the minor changes observed in the differential pattern are significant, or whether this pattern represents an exceptional case, we collected elemental abundances of metal-poor RPE stars from the literature, and calculated the mean scaled abundance patterns for $r$-I and $r$-II stars with [Fe/H] $< -2.0$. The choice of [Fe/H] $< -2.0$ is made to simply avoid the stars which may probably have the contribution of $s$-process in later Galactic chemical enrichment \citep{Hirai2019,     Tarumi.etal.2021}. For the mean abundance pattern calculation, we scaled individual abundance patterns to A(Eu)$ = 0$ by subtracting the Eu abundance of stars from their respective abundance patterns. The scaling to fixed A(Eu) value removes the large scatter in the mean abundance patterns due to different levels of $r$-process enhancement.

The top panel of Figure~\ref{fig:mean_diff_abundance_pattern} shows the mean scaled abundance patterns of $r$-I and $r$-II stars obtained from the literature along with the scaled-Solar abundance pattern. Near the second $r$-process peak, the mean scaled $r$-I and $r$-II patterns, displayed with cyan and magenta colors, respectively, follow the scaled-Solar $r$-process pattern shown with the black curve. This confirms the well-known universality of the main $r$-process pattern discussed in earlier studies \citep[see, e.g., the review from][]{sneden.etal.2008}. The mean $r$-I pattern deviates from the Solar $r$-process pattern, particularly towards lower atomic numbers. However, the $r$-II pattern closely follows a Solar $r$-process pattern. 

The bottom panel of Figure~\ref{fig:mean_diff_abundance_pattern} shows the differential pattern of mean scaled abundances from the literature in blue, along with the differential-abundance pattern from this study in yellow. The shaded blue color represents the 1$\sigma=\sqrt{\sigma_{r-I}^{2} + \sigma_{r-II}^{2}}$ dispersion. It is evident that the differential-abundance pattern from this study exhibits good agreement with the differential-abundance pattern obtained from the mean $r$-I and $r$-II patterns in the literature, suggesting that the differential pattern of HD~107752 with respect to CS~31082-0001 is typical to the general $r$-I and $r$-II population. The differential abundances of first-peak elements in this study exhibit a small excess compared to the mean differential pattern, which could be due to the contribution from a weak $r$-process site. Nevertheless, the differential-abundance pattern of the second $r$-process peak shows good consistency with the pattern obtained from the literature sample, suggesting a similar astrophysical site of origin, such as NSMs.

\section{Discussion}
\label{sec:discussion}

\subsection{Dependence of First r-process Peak Elements on Europium Enhancement}

\label{sec_large_sample}
\begin{figure*}
    \centering
    \includegraphics[width=\textwidth]{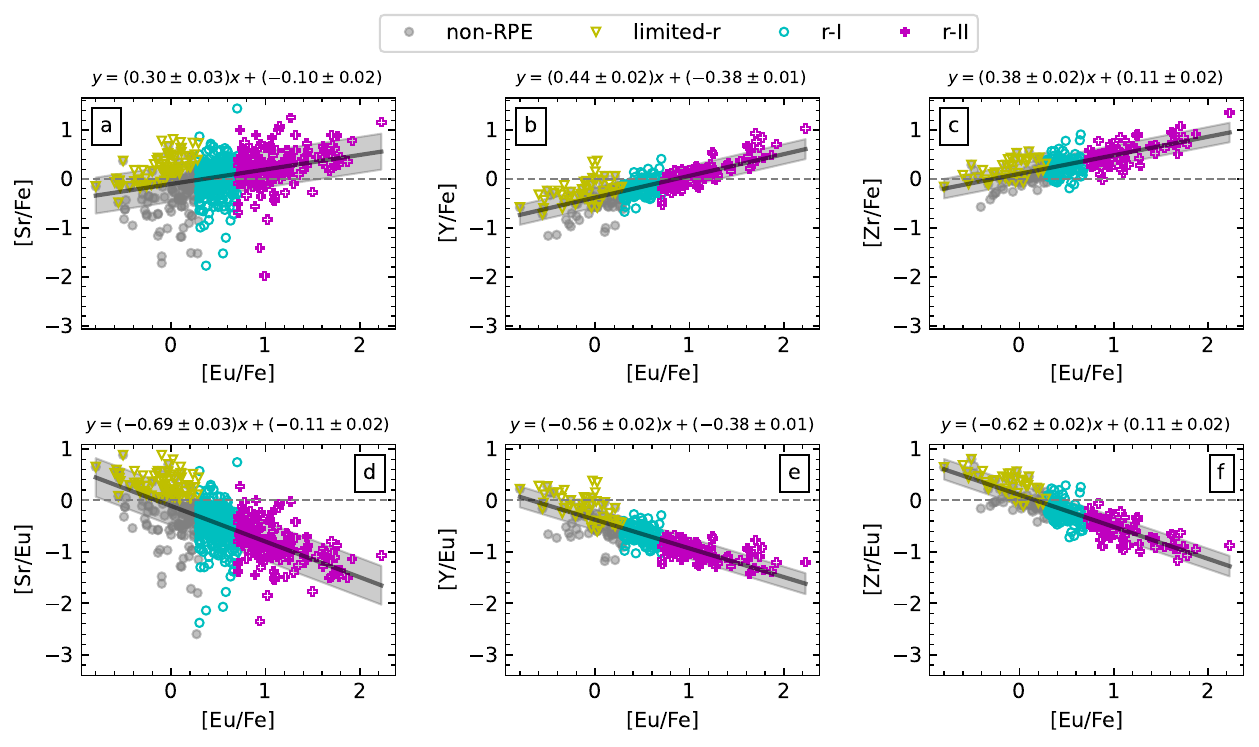}
    \caption{Top row: The evolution of first-peak $r$-process abundance ratios ([Sr/Fe], [Y/Fe], and [Zr/Fe]), as a function of [Eu/Fe].  Bottom row: The same elements are shown relative to the level of europium ([Sr/Eu], [Y/Eu], and [Zr/Eu]), as a function of [Eu/Fe]. The gray, yellow, cyan, and magenta colors represent non-RPE stars, limited-$r$ stars, $r$-I stars, and $r$-II stars, respectively. The black-solid lines are linear fits to the data, and the gray-shaded regions are the $1\sigma$ dispersions. The fitted equations are shown on the top of respective panels. The gray-dashed horizontal lines represent Solar values. For reference, see \cite{Saraf.etal.2023} and \cite{Saraf.etal.2023BINA}.}
    \label{fig:1st_2nd_peaks_rel}
\end{figure*}

To investigate the origin of the excess of first $r$-process peak elements in $r$-I stars, we have explored the relationship between [Sr/Fe], [Y/Fe], and [Zr/Fe] abundance ratios, as functions of their Eu enhancement, from our large compiled sample of metal-poor stars \citep[see][for data compilation]{Saraf.etal.2023}. The top row of Figure~\ref{fig:1st_2nd_peaks_rel} shows the evolution of these first-peak elements, as a function of [Eu/Fe], while in the bottom row, we illustrate the evolution of these elements relative to Eu, as a function of [Eu/Fe] enhancement. The gray, yellow, cyan, and magenta colors indicate non-RPE stars, limited-$r$, $r$-I and $r$-II stars, respectively. The black-solid lines represent linear fits to the sample, and the gray-shaded regions are their $1\sigma$ dispersion. The fitted equations are given on the top of the respective panel. For reference, the gray-dashed horizontal line indicates the Solar values. The top row of Figure~\ref{fig:1st_2nd_peaks_rel} suggests that the first $r$-process peak elements are produced concurrently with Eu.  However, the bottom row does not exhibit constant values of [Sr/Eu], [Y/Eu] and [Zr/Eu] with respect to the [Eu/Fe], as one would expect for a simple case of a single astrophysical site. The continuous decreasing trend between [Sr/Eu], [Y/Eu] and [Zr/Eu] with respect to the overall $r$-process enhancement possibly indicates a connection between the production sites of $r$-I and $r$-II stars.

\cite{Siqueira-Mello.etal.2014} found two distinct trends for the abundance ratios of  [Sr/Ba], [Y/Ba], and [Zr/Ba] as functions of [Ba/Fe]. They found linearly decreasing ratios with increasing [Ba/Fe] for [Ba/Fe] $> -1.5$ and constant ratios for [Ba/Fe] $< -1.5$, leading them to conclude that at least some fraction of the first $r$-process peak elements is produced independent of the second-peak elements. However, their study was based on a relatively small sample. In addition, most of the objects in their sample for [Ba/Fe] $< -1.5$ had only an upper limit. Based on our new compilation of additional samples (see Figure~\ref{fig:1st_2nd_peaks_rel_with_Ba} in Appendix~\ref{app:1st_2nd_peaks_rel_with_Ba}), we do not find two distinct trends. Instead, we find continuous linear anti-correlations similar to that we obtain with respect to Eu enhancement in Figure~\ref{fig:1st_2nd_peaks_rel}.

\begin{figure}
    \centering
    \includegraphics[width=0.45\columnwidth]{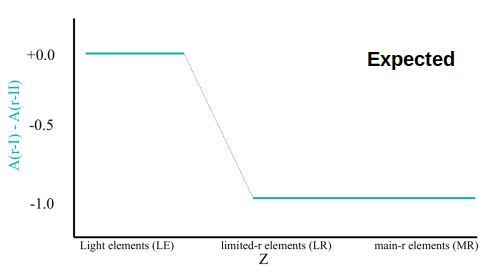}
    \includegraphics[width=0.45\columnwidth]{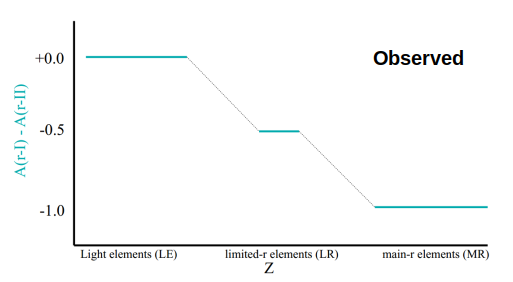}
    \caption{\textit{Left panel:} A schematic diagram illustrating the expected differential abundance pattern of HD~107752 relative to CS~31082-0001, if the neutron-capture elements are formed in the same kind of progenitor $r$-process sites. \textit{Right panel:} A schematic diagram explaining the expected differential-abundance pattern of HD~107752 with respect to CS~31082-0001 if the lighter and main $r$-process elements originate from different types of progenitor sites. This is similar to the observed differential abundance.}
    \label{fig:diff_illus}
\end{figure}
%

\begin{figure}
    \centering
    \includegraphics[width=0.6\columnwidth]{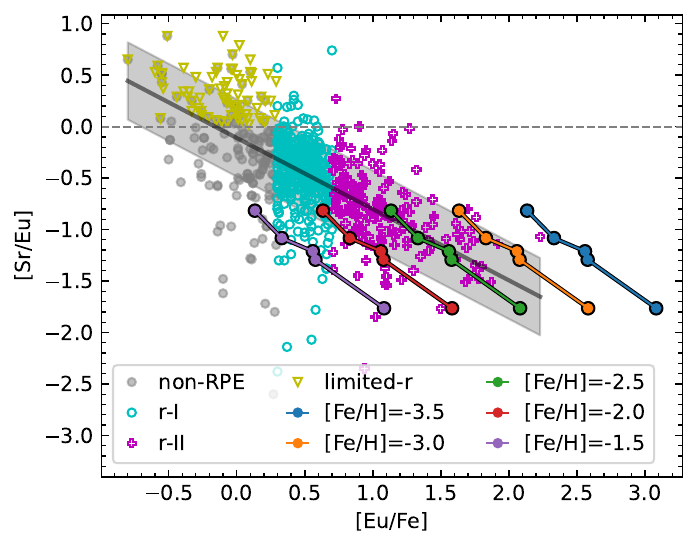}
    \caption{The evolution of [Sr/Eu], as a function of [Eu/Fe]. The gray, yellow, cyan, and magenta colors represent non-RPE stars, limited-$r$ stars, $r$-I stars, and $r$-II stars, respectively. The black-solid lines are linear fits to the data, and the gray-shaded regions are the $1\sigma$ dispersions. The gray-dashed horizontal lines represent Solar values. The predictions for mixing NSM ejecta in the ISM with metallicity of [Fe/H] = $-$3.5, $-$3.0, $-$2.5, $-$2.0, $-$1.5 are shown using blue, orange, green, red, and purple color curves, respectively. See text for details on mixing simulated NSM ejecta into the ISM.}
    \label{fig:data_and_model}
\end{figure}

\subsection{Necessity of Multiple Astrophysical Sites for r-I Stars}
The dilution of $r$-process ejecta from a single astrophysical site could form both $r$-I and $r$-II stars. Previous studies suggested that different degrees of dilution potentially produce scatters in $r$-process elements-to-Fe ratios \citep{vandeVoort2015, Hirai2015, Shen2015, Hirai2025, Hirai2017}. The only difference between $r$-I and $r$-II sub-classes of stars is the enhancement of neutron-capture elements, $r$-II being more enhanced than $r$-I. In this case, their abundance patterns should be similar, particularly for the main $r$-process peak elements. The ISM with [Eu/Fe] $>+0.7$ is the birthplace of $r$-II stars, and the ISM with $+0.3<$ [Eu/Fe] $\leq+0.7$ is the birthplace of $r$-I stars. According to this scenario, $r$-II stars are formed in clouds that receive a higher concentration of material from an $r$-process event, whereas $r$-I stars originate from the clouds where the material is more diluted. With a single astrophysical site, the ratio of neutron-capture elements would show constant values with respect to $r$-process enhancement.

The difficulty arises when considering the light $r$-process elements Sr, Y, and Zr. Our differential abundance analysis of the $r$-I star compared to the $r$-II star exhibits an excess of light $r$-process elements in the $r$-I star compared to the $r$-II star (Figure~\ref{fig:diff_abundance_pattern}). The top panel of Figure~\ref{fig:diff_illus} illustrates an expected differential-abundance pattern of HD 107752 with respect to CS 31082-0001 if the neutron-capture elements came from the same type of progenitor. In this case, the differential abundances of neutron-capture elements are expected to exhibit a plateau. However, the observed abundances in Figure~\ref{fig:diff_abundance_pattern} are similar to the illustration shown in the bottom panel of Figure~\ref{fig:diff_illus}, with three plaueaus at $-$0.01, $-$0.50, and $-$1.01 for the $\alpha$- and Fe-peak elements, lighter $r$-process elements, and the main $r$-process species, respectively. The abundances of CS 31082-0001 are ten times more enriched with main $r$-process than that of HD 107752, and the abundances of HD 107752 are three times (10$^{0.5}$) more enhanced with weak $r$-process elements than that of CS 31082-0001.

In order to better understand the production of the first and second $r$-process peak elements, we employ the simulations of NSMs reported in \citet{Fujibayashi.etal.2023}. In their comprehensive study of binary NSMs, \citet{Fujibayashi.etal.2023} investigate mass ejection and nucleosynthesis outcomes. They found that dynamical ejecta from asymmetric mergers are more neutron-rich, favoring the synthesis of heavier $r$-process nuclei, while post-merger disk ejecta -- characterized by milder neutron-richness -- contribute predominantly to lighter $r$-process elements. By combining both ejecta components, they could closely reproduce the Solar $r$-process pattern for different binary mass ratios. To calculate the [Eu/Fe] of the ISM polluted with the NSM ejecta, we assume a dwarf galaxy with a gas mass 10$^6$ $M_{\sun}$\footnote{This gas mass is the typical mass in the birthplace halo of RPE stars \citep{Hirai2022}.}; 75\% of this gas is hydrogen. We uniformly mix the NSM ejecta into this gas and calculate the [Eu/Fe] ratio for a range of ISM metallicities from [Fe/H] = $-3.5$ to $-1.5$ with a 0.5 dex step size. Figure~\ref{fig:data_and_model} shows the observed data of limited-$r$ (yellow down triangles), $r$-I (cyan circles), $r$-II (magenta crosses) stars over-plotted with the model values for five metallicies, [Fe/H] = $-3.5$ (blue curve), $-3.0$ (orange curve), $-2.5$ (green curve), $-2.0$ (red curve), and $-1.5$ (purple curve). At each metallicity, the different data points in the model represent NSM simulations with varying mass ratios. From low to high Eu abundance, the models correspond to SFHo130-140, SFHo135-135, SFHo120-150, SFHo125-145, and SFHo125-155 from \cite{Fujibayashi.etal.2023}, respectively. From inspection, the model reasonably reproduces the lower end of the [Sr/Eu] behavior, and naturally explains the [Sr/Eu] ratio and the overall $r$-process enhancement. However, the model does not reach higher values of [Sr/Eu] ratios, particularly for the $r$-I and $r$-II stars. Therefore, a different astrophysical site/condition that synthesizes only light $r$-process elements appears to be required.

As discussed above, $r$-I stars exhibit an excess of light $r$-process elements (Sr, Y, and Zr) as compared to $r$-II stars for a given overall $r$-process enhancement. \citet{Bandyopadhyay2024} also reported the excess of light $r$-process elements in $r$-I stars in their RPA data. These results suggest that the enrichment of the ISM with light $r$-process material, which is combined with dilution of $r$-process material, could explain the observed ratios of n-capture elements. 

The proposed candidate sites for light $r$-process elements are neutrino-driven winds in supernovae \citep{Arcones.Montes.2011}, electron-capture supernovae \citep{Wanajo.etal.2011}, and quark-deconfinement supernovae \citep{Fischer.etal.2020}. The $s$-process in fast rotating massive stars could also contribute to the Sr, Y, and Zr enrichment \citep{Prantzos2018}. Unfortunately, we do not presently have any observational evidence for the existence of these candidate sites. \citet{Hirai2019} have shown that the multiple astrophysical sites should contribute to the enrichment of Sr in their chemo-dynamical simulations. Distinct sites with different amounts of contribution from the disk and dynamical ejecta from an NSM event for the production of $r$-I and $r$-II stars could also be an interesting possibility to explore further. In NSMs, dynamical ejecta exhibit low electron fractions ($Y_e \lesssim 0.1$), indicating a high neutron-richness. These conditions strongly favor the synthesis of heavier $r$-process nuclei, typically with mass numbers $A \gtrsim 140$ \citep{Goriely.etal.2011, Fujibayashi.etal.2023}. Conversely, disk ejecta tends to have moderate electron fractions ($Y_e \approx 0.2$--$0.3$). This results in nucleosynthesis dominated by lighter $r$-process elements, with mass numbers primarily in the range $A \approx 90$--$130$ \citep{Rosswog.etal.2014, Wanajo.etal.2014, Fujibayashi.etal.2020}.

\section{Conclusion}
\label{sec:conclusion}
In this work, we have studied the differential-abundance patterns of a $r$-I star (HD~107752) with respect to a $r$-II star, CS~31082-0001, based on a strictly line-by-line abundance estimation of 16 light elements and 15 neutron-capture elements. The precision of the abundance determination in this work is better than 0.08 dex. This improved precision allows us to constrain the sites of the natal clouds from where these RPE stars formed. We note three distinct regions in differential-abundance patterns (see Figure~\ref{fig:diff_abundance_pattern}). Light elements exhibit differential abundances very close to zero, suggesting their similar origin in both the stars, e.g., CCSNe. However, differential abundances of the first and second $r$-process peak elements suggest that they are formed under different astrophysical conditions. Simple dilution of the $r$-process yield from a single site such as a NSM, collapsar, or magneto-rotational supernova, cannot explain the observed differential-abundance pattern.

We also report a clear pattern in the differential abundances of second $r$-process peak elements, in contrast to the universal main $r$-process pattern routinely claimed in previous studies of RPE stars. The mean differential-abundance pattern of previously known $r$-I and $r$-II stars with [Fe/H] $< -2.0$ further supports the significance and reliability of the reported differential pattern in the main $r$-process region. A more detailed differential-abundance analysis of a large sample of RPE stars may provide key information to understand $r$-process nucleosynthesis and its production sites.

In our study, we also compared observational trends of the [Sr/Eu] versus [Eu/Fe] ratios with NSM models. Our findings indicate that these models can reasonably explain the trends in [Sr/Eu] versus [Eu/Fe] for $r$-II stars, but they fail to extend the relation further to $r$-I and limited-$r$ stars. Notably, the model corresponding to a metallicity of [Fe/H] $\approx -$2.5 aligns well with a straight-line fit to the observational data, suggesting a partial alignment, but also highlighting discrepancies that warrant further investigation.

We discuss the importance of multiple sites in the formation of $r$-I stars. Our analysis conclude that neither the dilution of ejecta nor pre-enriched ISM could adequately explain the observed differential abundance pattern and light-to-heavy $r$-process element ratio trends. Instead, it may be worth exploring the possibility that varying contributions from disk and dynamical ejecta across different NSM sites play a role in the formation of $r$-I and $r$-II stars. This model can naturally explain the strong correlation among $r$-process rich stars regarding the ratio of light-to-heavy neutron capture elements as a function of $r$-process enhancement. 


\begin{acknowledgments}
The author thank the ESO archive for compiling all the suitable data for the public. This work makes use of data from the European Space Agency (ESA) mission {\it Gaia} (\url{https://www.cosmos.esa.int/gaia}), processed by the {\it Gaia} Data Processing and Analysis Consortium (DPAC, \url{https://www.cosmos.esa.int/web/gaia/dpac/consortium}). Funding for the DPAC has been provided by national institutions, in particular the institutions participating in the {\it Gaia} Multilateral Agreement. The work at PRL is supported by the Department of Space. PS and TS thank Prof. Shinya Wanajo for his helpful discussions during the process. PS is grateful to Prof. Shashikiran Ganesh for his support and for the freedom to continue this work at PRL. T.C.B. acknowledges partial support from grants PHY 14-30152; Physics Frontier Center/JINA Center for the Evolution of the Elements (JINA-CEE), and OISE-1927130: The International Research Network for Nuclear Astrophysics (IReNA), awarded by the US National Science Foundation. Y. H. was supported in part by JSPS KAKENHI Grant Numbers JP22KJ0157, JP25H00664, JP25K01046, MEXT as ``Program for Promoting Researches on the Supercomputer Fugaku" (Structure and Evolution of the Universe Unraveled by Fusion of Simulation and AI; Grant Number JPMXP1020230406), and JICFuS. CAP is thankful for funding from the Spanish government through grants AYA2014-56359-P, AYA2017-86389-P and PID2020-117493GB-100. 
\end{acknowledgments}

%
\vspace{5mm}
\facilities{VLT(UVES)}

\software{ARES \citep{ARES.2007, ARES.2015},
    Astropy \citep{astropy:2013, astropy:2018,  astropy:2022},
    ATLAS9 \citep{ATLAS.1993, ATLAS.2003},
    Gala \citep{gala_code.2017},
    IRAF,
    Matplotlib \citep{Matplotlib.2007},
    NumPy \citep{Numpy.2020},
    Pandas \citep{Pandass.paper.2010, Pandas.software.2020},
    TURBOSPECTRUM \citep{Turbospectrum.1998, Turbospectrum.2012}
    }


\appendix

\section{Line list}
\label{app:linelist}
In this work, we compile the atomic and molecular line list from \cite{Heiter.etal.2015} and \cite{Heiter.2020}. Table~\ref{app:linelist} shows our complete line list of elements identified for differential analysis in both the stars. We have tabulated the names of elements along with their ionization states, wavelengths, lower excitation potentials (LEP), $\log gf$ values, and equivalent widths for the stars CS~31082-0001 and HD~107752.

\begin{longtable}{lccrrr}
	\caption{Information on the adopted line list. References are available in \cite{Heiter.etal.2015} and \cite{Heiter.2020}}
	\label{tab:line_list_of_other_elements}\\
	\hline
	Element & Wavelength ({\AA}) & LEP (eV) & $\log gf$ & EqW (m{\AA}) & EqW (m{\AA})\\
     & & & & CS~31082-0001 & HD~107752\\
	\hline
Fe I & 4009.713 & 2.223 & $-$1.252 & 53.5 & 55.9\\
Fe I & 4014.531 & 3.047 & $-$0.587 & 41.8 & 52.7\\
Fe I & 4134.677 & 2.831 & $-$0.649 & 44.8 & 41.8\\
Fe I & 4143.415 & 3.047 & $-$0.204 & 44.8 & 53.7\\
Fe I & 4175.636 & 2.845 & $-$0.827 & 47.8 & 46.4\\
Fe I & 4181.755 & 2.831 & $-$0.371 & 58.1 & 60.8\\
Fe I & 4182.383 & 3.017 & $-$1.180 & 17.6 & 17.4\\
Fe I & 4191.430 & 2.469 & $-$0.666 & 62.0 & 67.6\\
Fe I & 4199.095 & 3.047 & 0.155 & 66.3 & 69.5\\
Fe I & 4222.213 & 2.449 & $-$0.967 & 56.5 & 58.5\\
Fe I & 4233.602 & 2.482 & $-$0.604 & 67.1 & 69.0\\
Fe I & 4250.119 & 2.469 & $-$0.405 & 73.7 & 79.7\\
Fe I & 4260.474 & 2.399 & 0.109 & 96.2 & 101.5\\
Fe I & 4271.153 & 2.449 & $-$0.349 & 82.5 & 88.3\\
Fe I & 4282.403 & 2.176 & $-$0.779 & 78.0 & 75.3\\
Fe I & 4427.310 & 0.052 & $-$2.924 & 90.9 & 107.5\\
Fe I & 4433.218 & 3.654 & $-$0.700 & 9.0 & 12.6\\
Fe I & 4442.339 & 2.198 & $-$1.255 & 62.4 & 64.4\\
Fe I & 4447.717 & 2.223 & $-$1.342 & 49.6 & 55.9\\
Fe I & 4459.117 & 2.176 & $-$1.279 & 66.2 & 67.4\\
Fe I & 4466.551 & 2.831 & $-$0.600 & 59.8 & 67.0\\
Fe I & 4494.563 & 2.198 & $-$1.136 & 60.3 & 69.1\\
Fe I & 4632.911 & 1.608 & $-$2.913 & 15.8 & 20.5\\
Fe I & 4710.283 & 3.018 & $-$1.612 & 6.6 & 8.5\\
Fe I & 4733.591 & 1.485 & $-$2.988 & 19.3 & 24.0\\
Fe I & 4736.772 & 3.211 & $-$0.752 & 25.5 & 35.0\\
Fe I & 4871.317 & 2.865 & $-$0.363 & 61.1 & 65.0\\
Fe I & 4872.138 & 2.882 & $-$0.567 & 52.9 & 53.6\\
Fe I & 4890.754 & 2.875 & $-$0.394 & 65.9 & 62.1\\
Fe I & 4891.492 & 2.851 & $-$0.111 & 69.3 & 75.8\\
Fe I & 4903.310 & 2.882 & $-$0.926 & 30.2 & 36.1\\
Fe I & 4982.537 & 4.283 & $-$3.609 & 15.2 & 19.4\\
Fe I & 4994.130 & 0.915 & $-$3.002 & 48.4 & 55.8\\
Fe I & 5001.863 & 3.881 & $-$0.010 & 20.1 & 25.9\\
Fe I & 5006.117 & 2.832 & $-$0.638 & 50.8 & 52.6\\
Fe I & 5014.942 & 3.943 & $-$0.303 & 14.4 & 18.3\\
Fe I & 5039.251 & 3.368 & $-$1.573 & 7.1 & 8.8\\
Fe I & 5049.820 & 2.279 & $-$1.349 & 48.7 & 55.9\\
Fe I & 5051.634 & 0.915 & $-$2.795 & 62.0 & 72.7\\
Fe I & 5079.740 & 0.990 & $-$3.233 & 37.3 & 40.5\\
Fe I & 5125.112 & 4.220 & $-$0.140 & 11.4 & 15.9\\
Fe I & 5133.688 & 4.178 & 0.140 & 22.9 & 27.2\\
Fe I & 5150.839 & 0.990 & $-$3.020 & 37.5 & 45.5\\
Fe I & 5162.273 & 4.178 & 0.020 & 19.9 & 22.2\\
Fe I & 5192.343 & 2.998 & $-$0.421 & 44.1 & 51.6\\
Fe I & 5194.941 & 1.557 & $-$2.021 & 63.1 & 69.8\\
Fe I & 5198.711 & 2.223 & $-$2.113 & 21.5 & 21.7\\
Fe I & 5202.335 & 2.176 & $-$1.838 & 38.1 & 40.6\\
Fe I & 5215.180 & 3.266 & $-$0.871 & 16.9 & 21.7\\
Fe I & 5216.274 & 1.608 & $-$2.082 & 55.2 & 60.7\\
Fe I & 5217.389 & 3.211 & $-$1.116 & 13.4 & 19.6\\
Fe I & 5232.940 & 2.940 & $-$0.076 & 71.6 & 75.5\\
Fe I & 5247.050 & 0.087 & $-$4.975 & 9.8 & 14.5\\
Fe I & 5250.209 & 0.121 & $-$4.918 & 12.5 & 13.2\\
Fe I & 5250.646 & 2.198 & $-$2.180 & 27.3 & 26.0\\
Fe I & 5266.555 & 2.998 & $-$0.386 & 52.4 & 55.9\\
Fe I & 5281.790 & 3.038 & $-$0.833 & 29.3 & 33.9\\
Fe I & 5324.179 & 3.211 & $-$0.103 & 53.2 & 59.9\\
Fe I & 5332.899 & 1.557 & $-$2.777 & 22.0 & 26.1\\
Fe I & 5339.929 & 3.266 & $-$0.684 & 24.7 & 31.4\\
Fe I & 5367.466 & 4.415 & 0.444 & 16.3 & 20.7\\
Fe I & 5369.961 & 4.371 & 0.536 & 18.3 & 23.7\\
Fe I & 5383.369 & 4.312 & 0.632 & 29.8 & 30.4\\
Fe I & 5415.199 & 4.386 & 0.643 & 22.1 & 27.6\\
Fe I & 5424.068 & 4.320 & 0.520 & 29.9 & 32.8\\
Fe I & 5434.524 & 1.011 & $-$2.119 & 96.1 & 107.2\\
Fe I & 5445.042 & 4.386 & $-$0.020 & 10.8 & 13.1\\
Fe I & 5501.465 & 0.958 & $-$3.046 & 53.1 & 58.4\\
Fe I & 5506.779 & 0.990 & $-$2.793 & 61.5 & 71.4\\
Fe I & 5572.842 & 3.396 & $-$0.275 & 36.2 & 38.6\\
Fe I & 5576.089 & 3.430 & $-$0.900 & 15.2 & 19.0\\
Fe I & 6065.482 & 2.608 & $-$1.470 & 27.4 & 28.1\\
Fe I & 6136.615 & 2.453 & $-$1.405 & 44.4 & 43.8\\
Fe I & 6137.691 & 2.588 & $-$1.375 & 33.6 & 36.6\\
Fe I & 6230.722 & 2.559 & $-$1.279 & 45.2 & 47.1\\
Fe I & 6252.555 & 2.404 & $-$1.727 & 33.5 & 35.9\\
Fe I & 6393.600 & 2.433 & $-$1.504 & 39.2 & 40.0\\
Fe I & 6421.350 & 2.279 & $-$2.020 & 26.4 & 29.5\\
Fe I & 6430.845 & 2.176 & $-$1.976 & 31.3 & 36.6\\
Fe I & 6593.870 & 2.433 & $-$2.394 & 9.5 & 10.9\\
Fe II & 4489.183 & 2.828 & $-$2.971 & 14.4 & 16.7\\
Fe II & 4491.405 & 2.856 & $-$2.756 & 19.5 & 22.1\\
Fe II & 4508.288 & 2.856 & $-$2.349 & 33.7 & 38.2\\
Fe II & 4923.927 & 2.891 & $-$1.260 & 82.6 & 89.9\\
Fe II & 5018.440 & 2.891 & $-$1.100 & 91.8 & 99.8\\
Fe II & 5276.002 & 3.199 & $-$2.213 & 31.6 & 36.7\\
Fe II & 5362.869 & 3.199 & $-$2.570 & 15.4 & 18.4\\
Ca I & 6493.781 & 2.521 & $-$0.109 & 28.2 & 25.4\\
Ca I & 6449.808 & 2.521 & $-$0.502 & 13.1 & 13.2\\
Ca I & 6439.075 & 2.526 & 0.390 & 50.1 & 49.6\\
Ca I & 6102.723 & 1.879 & $-$0.793 & 29.6 & 27.3\\
Ca I & 5588.749 & 2.526 & 0.358 & 43.4 & 40.9\\
Ca I & 5590.114 & 2.521 & $-$0.571 & 10.2 & 9.2\\
Ca I & 4425.437 & 1.879 & $-$0.358 & 47.3 & 39.8\\
Co I & 4121.311 & 0.923 & $-$0.320 & 73.0 & 69.5\\
Cr I & 4646.162 & 1.030 & $-$0.700 & 22.5 & 26.2\\
Cr I & 5345.796 & 1.004 & $-$0.950 & 20.5 & 21.1\\
Cr I & 5348.315 & 1.004 & $-$1.210 & 11.2 & 11.4\\
Cr I & 5298.272 & 0.983 & $-$1.140 & 13.8 & 16.7\\
Cr II & 4824.127 & 3.871 & $-$0.980 & 17.0 & 11.2\\
Mg I & 5183.604 & 2.717 & $-$0.239 & 201.0 & 185.4\\
Mg I & 5172.684 & 2.712 & $-$0.450 & 173.7 & 169.2\\
Mg I & 4702.991 & 4.346 & $-$0.440 & 62.1 & 50.8\\
Mn I & 4823.524 & 2.319 & 0.136 & 13.1 & 10.9\\
Mn I & 4783.427 & 2.298 & 0.044 & 11.9 & 8.7\\
Mn I & 4033.062 & 0.000 & $-$0.618 & 95.5 & 90.6\\
Mn I & 4041.355 & 2.114 & 0.285 & 22.7 & 20.7\\
Na I & 5889.951 & 0.000 & 0.108 & 154.4 & 154.4\\
Ni I & 5080.528 & 3.655 & 0.330 & 13.7 & 8.74\\
Ni I & 5081.107 & 3.847 & 0.462 & 10.8 & 6.7\\
Ni I & 4904.407 & 3.542 & $-$0.016 & 6.9 & 5.0\\
Ni I & 4714.408 & 3.380 & 0.260 & 20.4 & 15.5\\
Sc II & 5526.790 & 1.768 & 0.024 & 22.7 & 16.7\\
Sc II & 4670.407 & 1.357 & $-$0.576 & 14.7 & 14.7\\
Sc II & 4324.996 & 0.595 & $-$0.442 & 63.8 & 61.7\\
Sc II & 4415.557 & 0.595 & $-$0.668 & 50.3 & 42.1\\
Si I & 4102.936 & 1.909 & $-$2.827 & 64.0 & 65.0\\
Ti I & 5210.385 & 0.048 & $-$0.827 & 29.6 & 24.5\\
Ti I & 5173.743 & 0.000 & $-$1.062 & 19.2 & 17.2\\
Ti I & 5036.464 & 1.443 & 0.130 & 6.4 & 4.6\\
Ti I & 5016.161 & 0.848 & $-$0.518 & 8.1 & 7.2\\
Ti I & 5022.868 & 0.826 & $-$0.378 & 12.6 & 9.0\\
Ti I & 5024.844 & 0.818 & $-$0.546 & 9.3 & 6.6\\
Ti I & 4981.731 & 0.848 & 0.560 & 47.6 & 42.1\\
Ti I & 4656.469 & 0.000 & $-$1.345 & 11.9 & 10.8\\
Ti II & 5418.751 & 1.582 & $-$2.110 & 16.0 & 12.8\\
Ti II & 5381.015 & 1.566 & $-$1.970 & 23.3 & 18.6\\
Ti II & 5185.902 & 1.893 & $-$1.490 & 22.4 & 20.4\\
Ti II & 5154.068 & 1.566 & $-$1.750 & 29.5 & 25.6\\
Ti II & 5013.677 & 1.582 & $-$2.190 & 14.0 & 10.7\\
Ti II & 4779.985 & 2.048 & $-$1.248 & 21.6 & 18.0\\
Ti II & 4501.270 & 1.116 & $-$0.770 & 99.1 & 97.7\\
Ti II & 4443.794 & 1.080 & $-$0.720 & 102.7 & 101.5\\
Ti II & 4444.555 & 1.116 & $-$2.240 & 31.6 & 29.5\\
Ti II & 4411.925 & 1.224 & $-$2.520 & 13.8 & 9.7\\
Ti II & 4394.051 & 1.221 & $-$1.780 & 47.3 & 47.7\\
Ti II & 4395.839 & 1.243 & $-$1.930 & 39.1 & 32.8\\
V I & 4406.633 & 0.301 & $-$0.190 & 14.5 & 7.0\\
V II & 4023.378 & 1.805 & $-$0.689 & 15.8 & 12.8\\
V II & 4036.777 & 1.476 & $-$1.594 & 5.1 & 5.0\\
Zn I & 4810.528 & 4.078 & $-$0.160 & 12.8 & 11.8\\
    \hline
\end{longtable}
 
\section{Comparison of Absorption Lines}
\label{app:line_comparison}
\begin{figure*}
    \centering
    \includegraphics[width=0.9\textwidth]{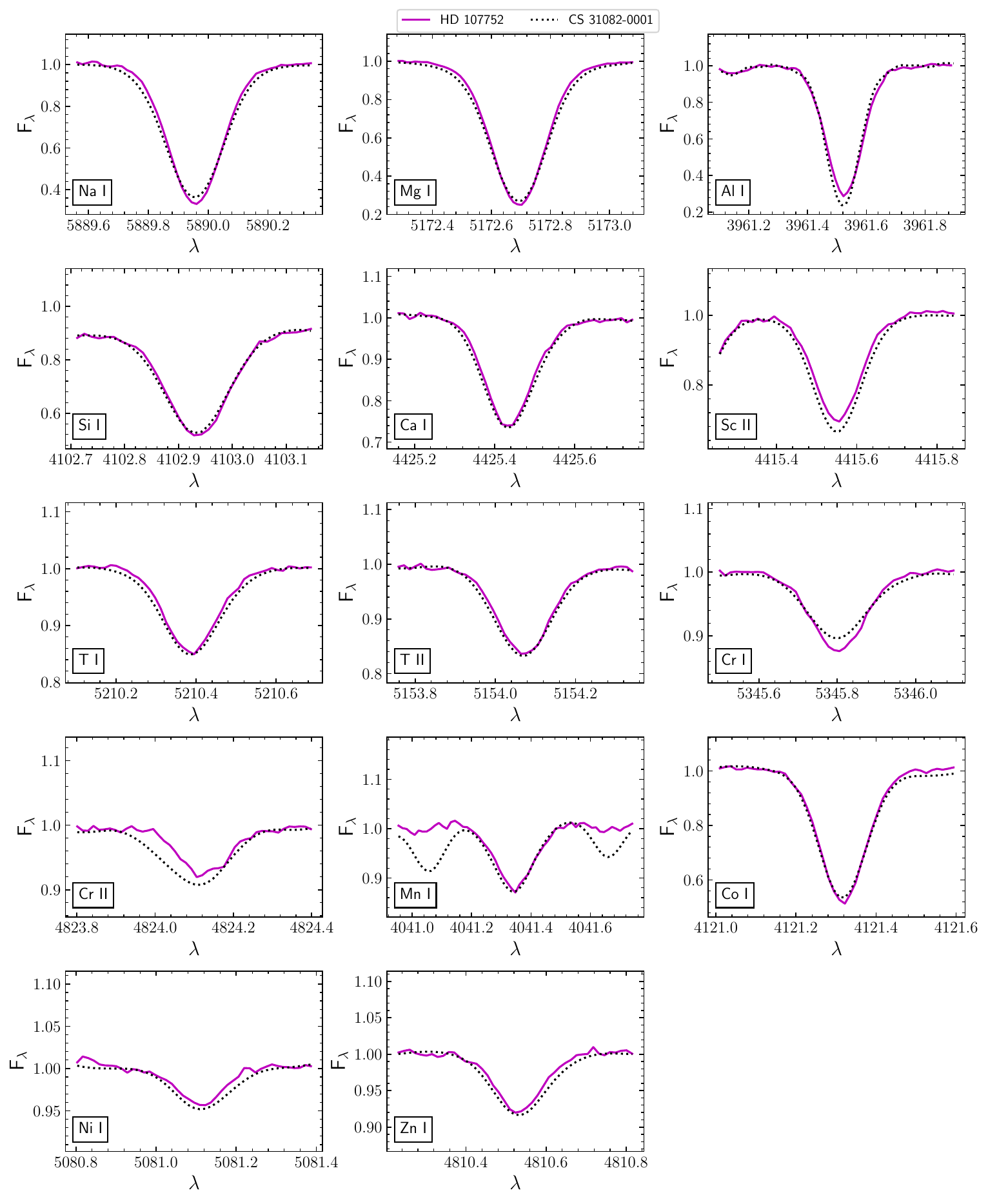}
    \caption{Comparison over-plots of different regions corresponding to the listed non neutron-capture elements of CS 31082-001 (dotted black curves) and HD 107752 (solid magenta curves)}
    \label{fig:light_lines_comparison}
\end{figure*}

\begin{figure*}
    \centering
    \includegraphics[width=\textwidth]{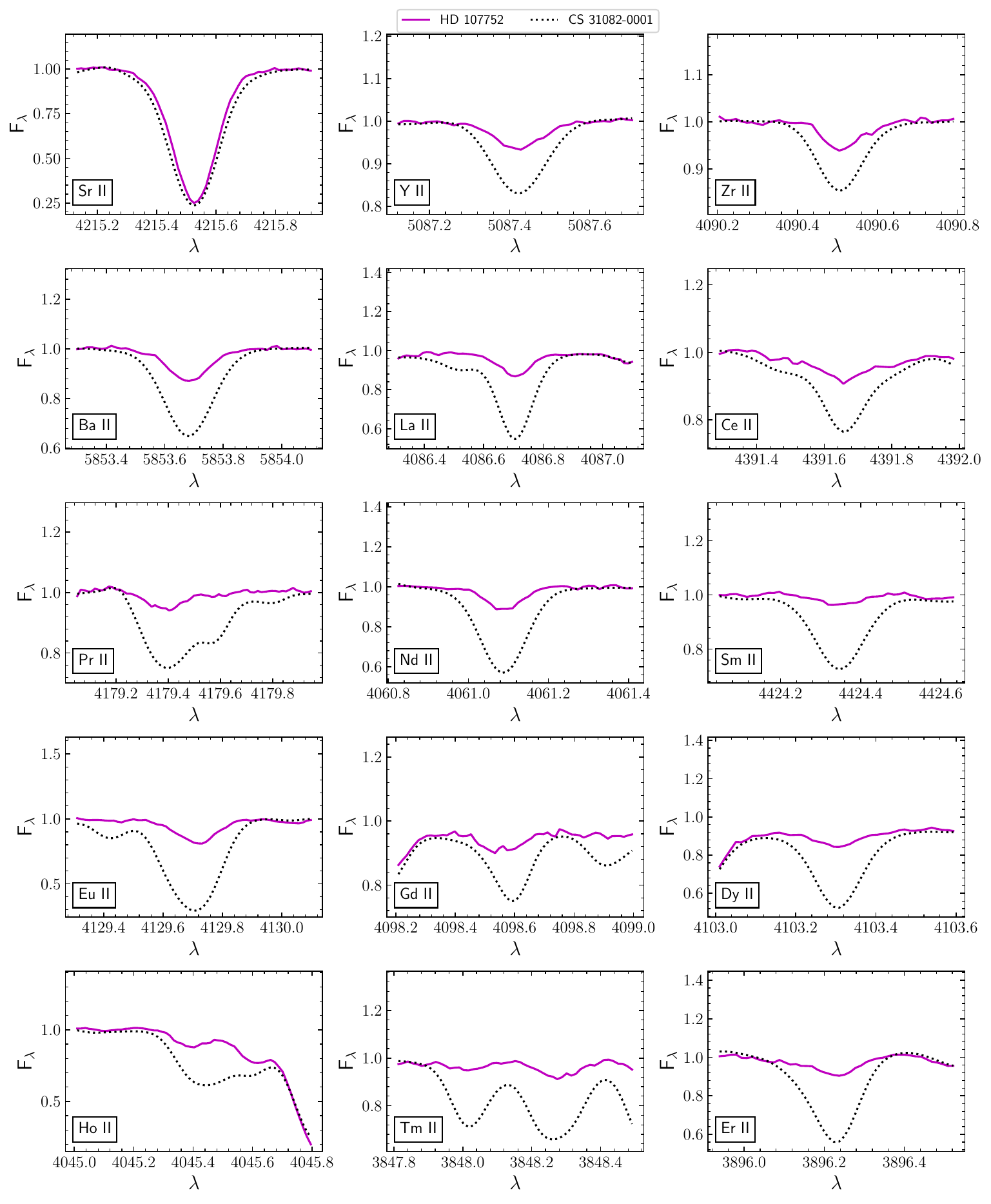}
    \caption{Comparison over-plot of different regions corresponding to the listed neutron-capture elements of CS 31082-001 (dotted black curves) and HD 107752 (solid magenta curves).}
    \label{fig:heavy_lines_comparison}
\end{figure*}

For a visualization of the differences in spectral lines between our two stars, we compare one absorption line for each of the non neutron-capture elements in Figure~\ref{fig:light_lines_comparison} and neutron-capture elements in Figure~\ref{fig:heavy_lines_comparison}. In each Figure, the dotted black curves are for CS 31082-0001 and the solid magenta curves are for HD 107752. The line profiles of non neutron-capture elements are very similar in both the stars, indicating their similar abundances in both the stars. In contrast, the neutron-capture elements show substantially stronger lines in CS 31082-001 as compared to HD 107752.

\newpage
\section{Dependence of first r-process peak elements on barium enhancement}
\label{app:1st_2nd_peaks_rel_with_Ba}
\begin{figure*}
    \centering
    \includegraphics[width=\textwidth]{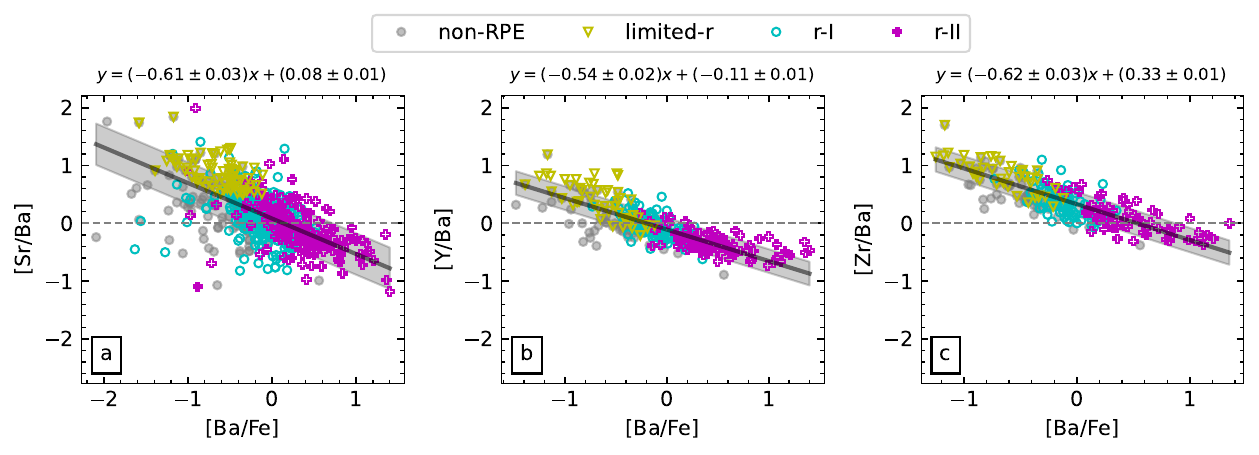}
\caption{The evolution of first-peak $r$-process abundance ratios ([Sr/Ba], [Y/Ba], and [Zr/Ba]), as a function of [Ba/Fe]. The gray, yellow, cyan, and magenta colors represent non-RPE stars, limited-$r$ stars, $r$-I stars, and $r$-II stars, respectively. The black-solid lines are linear fits to the data, and the gray-shaded regions are the $1\sigma$ dispersions. The fitted equations are shown on the top of respective panels. The gray-dashed horizontal lines represent Solar values. For reference, see \cite{Saraf.etal.2023} and \cite{Saraf.etal.2023BINA}.}
    \label{fig:1st_2nd_peaks_rel_with_Ba}
\end{figure*}

In Figure~\ref{fig:1st_2nd_peaks_rel_with_Ba}, we have shown the [Sr/Ba], [Y/Ba], and [Zr/Ba] as a function of barium enhancement. The gray circles, yellow down triangles, cyan circles, and magenta crosses, respectively, represent the non-RPE, limited-$r$, $r$-I, and $r$-II stars. Likewise with europium enhancement, they show a clear anti-correlation with barium enhancement.

\bibliography{references}{}
\bibliographystyle{aasjournalv7}



\end{document}